# Improving the recombination estimation method of Padhukasahasram et al 2006

Badri Padhukasahasram

Section on Ecology and Evolution, University of California Davis

Address for correspondence:
4341 Genome Center GBSF Building
451 Health Sciences Drive
University of California Davis, Davis, California, U.S.A. 95616

Email: bpadhuka@ucdavis.edu

**Abstract**
The accuracy of the recombination estimation method of Padhukasahasram et al. 2006 can be improved by including additional informative summary statistics in the rejection scheme and by simulating datasets under a fixed segregating sites model. A C++ program that outputs these summary statistics is freely available from my website.

We recently proposed a method for jointly estimating the population crossing-over and gene-conversion parameters from single-nucleotide polymorphism (SNP) data (Padhukasahasram et al. Genetics 2006). Here, I describe minor modifications to this method by including additional informative summary statistics and simulating datasets under a fixed segregating model where the positions of segregating sites are fixed to the same as observed in a real dataset and the minor allele frequencies at each position are roughly similar to what is observed in real data. In particular, I include the number of distinct haplotypes ($H$) in the rejection-sampling scheme, which is informative about the recombination rate (see Wall 2000) and increases monotonically as the rate increases. I also utilize, new, short-range versions of this summary statistic, which are defined as the arithmetic mean of the number of distinct haplotypes for short overlapping windows (e.g. 5 kb or 10 kb) along the sequence. Simulations showed that when crossing-over and conversion rates are assumed to be uniform along the sequence, the new implementation is more accurate than the original method.

I show comparisons with other current methods for estimating crossing-over rates alone (for parameters tested in Smith and Fearnhead 2005) in Table 1. These comparisons suggest that for the crossing-over estimation problem the accuracy obtained using my summaries is close to the best of the currently available methods.

Including the bounds on the minimum number of recombination events (Myers and Griffiths 2003, Song, Wu and Gusfield 2005 or Bafna and Bansal 2005) in the rejection scheme and smoothing the likelihood curves (for crossing-over estimation) or likelihood surfaces (for joint estimation) can further improve the relative efficiency of the method. I expect that the improved version of my method will perform well in other parameters too, but note that there are no theoretical guarantees in untested parameters.

A C++ program that implements the fixed segregating sites model and outputs these summary statistics, is freely available at my website: http:// badri-populationgeneticsimulators.blogspot.com

**Table 1: Comparison of methods for estimating crossing-over rates alone**

| Length (kb) | ρ per kb | Maxhap (Hudson 2001) | | SequenceLD (Fearnhead and Donnelly 2002) | | Rholike (Li and Stephens 2003) | | Padhukasahasram et al 2006 | |
|---|---|---|---|---|---|---|---|---|---|
| | | RMSE | g | RMSE | g | RMSE | g | RMSE | g |
| 2 | 1.00 | 1.70 | 0.39 | 1.41 | 0.38 | 1.26 | 0.29 | 1.632 | 0.31 |
| 2 | 4.00 | 1.43 | 0.57 | 1.29 | 0.62 | 1.20 | 0.44 | 1.310 | 0.62 |
| 2 | 16.00 | 0.68 | 0.60 | 0.60 | 0.70 | 0.66 | 0.59 | 0.669 | 0.82 |
| 10 | 0.25 | 0.81 | 0.64 | 1.02 | 0.43 | 1.05 | 0.56 | 0.898 | 0.68 |
| 10 | 1.00 | 0.58 | 0.86 | 0.57 | 0.80 | 0.47 | 0.90 | 0.527 | 0.89 |
| 10 | 4.00 | 0.37 | 0.93 | 0.32 | 0.95 | 0.23 | 0.98 | 0.423 | 0.96 |
| 25 | 1.00 | 0.36 | 0.73 | 0.33 | 0.71 | 0.31 | 0.85 | 0.356 | 0.76 |

Summary statistics that were used for estimating crossing-over rates are:
i)  Fraction of SNP pairs with $D' < 1$ (with 30% acceptance error).
ii) Fraction of ordered triplets $A$, $B$, $C$ with $D'(AB) < 0.5$ and $D'(BC) < 0.5$ (with 30% acceptance error).
iii) The number of distinct haplotypes $H$ (with 15% acceptance error).

Accuracy for the summary statistics method are based on 100 simulated datasets for all the values. For other methods, the numbers are same as in Table 1 in Smith and Fearnhead 2005. *RMSE* denotes the root mean square relative error for the estimates and *g* denotes the fraction of the datasets for which estimates are within a factor of either 1.5 (for 25 kb sequence) or 2.0 (for the rest) of the true values.